\documentclass[twocolumn,showpacs,amsmath,amssymb,prb,superscriptaddress]{revtex4}

\usepackage{graphicx}
\usepackage{bm}

\begin{document}

\title{Induced superconductivity in 2D electronic systems}

\author{ N.B.\ Kopnin }
\affiliation{ Low Temperature Laboratory, Aalto University, P.O.
Box 15100, 00076 Aalto, Finland} \affiliation{ L.~D.~Landau
Institute for Theoretical Physics, 117940 Moscow, Russia}
\author{A.S.\ Melnikov} \affiliation{Institute for Physics
of Microstructures, Russian Academy of Sciences, 603950 Nyzhny
Novgorod, GSP-105, Russia }

\date{\today}

\begin{abstract}
The approach applicable for spatially inhomogeneous and
time-dependent problems associated with the induced
superconductivity in low dimensional electronic systems is
developed. This approach is based on the Fano--Anderson model
which describes the decay of a resonance state coupled to a
continuum. We consider two types of junctions made of a ballistic
2D electron gas placed in a tunnel finite-length contact with a
bulk superconducting leads. We calculate the spectrum of the bound
states, supercurrent, and the current-voltage curve which show a
rich structure due to the presence of induced gap and dimensional
quantization.
\end{abstract}
\pacs{74.45.+c, 74.50.+r, 74.78.-w}

\maketitle

\section{Introduction}

Recent progress in studies of  transport in graphene that followed
the seminal work of Ref. \cite{Novoselov05} has boosted the
interest in properties of contacts between the graphene sheets and
various types of electrodes attached to them. Of special interest
are contacts between graphene and superconducting electrodes due
to the specific nature of the Andreev reflection in graphene
\cite{Beenakker06}. In practical devices, such type of contacts
has the form of a superconducting lead placed on top of the
graphene layer which partially overlaps with the lead. Following
Ref. \cite{Beenakker06}, the model most commonly used for
description of such contacts assumes that the superconductor
simply introduces certain pairing potential in the part of layer
which is immediately under the superconductor as well as shifts
its Fermi level away from the Dirac point. As a result, the
contact is treated as being formed between the usual normal
graphene layer and such piece of graphene where both the induced
pairing potential and the high doping level are present. Various
modifications of this model have been extensively used for
studying transport properties of graphene contacts
\cite{Tworzylo06,CuevasYeyati06,LevyYeyati08,LevyYeyati09,BlackShafferDoniach08,Linder09,Rainis09}.

Though this model is a significant step forward in understanding
the induced superconductivity in graphene, it still oversimplifies
the proximity effects which the bulk superconductor has on the
underlying normal sheet. To study these effects more carefully one
can look at other models used for contacts between superconductors
and low-dimensional electronic systems. For example, in Refs.
\cite{AVolkov95,Fagas-etal-05} the proximity effect in a
two-dimensional (2D) electron gas was modelled by a uniform plane
contact between a bulk superconductor and a thin normal conducting
layer. Such model correctly catches the main physics of the
proximity effect in spatially homogeneous structures in
equilibrium. In particular, it accounts for the induced energy gap
that depends on the contact resistance. However, it cannot
describe spatially-dependent problems in 2D systems nor can it be
easily extended to time-dependent or non-equilibrium phenomena.

In the present paper we develop an approach which is suitable for
spatially inhomogeneous and/or time-dependent problems associated
with the induced superconductivity in 2D systems placed in a
contact with a bulk superconductor. Our model is similar to that
used in Refs. \cite{Shiba,ArseevVolkov91} for impurities in a
superconductor and is based on the so-called Fano--Anderson model
which describes the decay of a localized state coupled to a
continuum \cite{FanoAnderson}. Our model can be applied to various
2D electronic systems, including simple 2D gas, graphene layer,
etc. In this paper we consider two particular examples of
junctions made of a 2D ballistic electron gas placed under the
superconducting electrodes. The application of our model to the
induced superconductivity in graphene will be considered
elsewhere.

The paper is organized as follows. In the next Section we describe
our  proximity model in its general formulation suitable for
various applications. The particular example of a junction between
two proximity-induced superconductors is considered in Section
\ref{sec-SDS}. We solve equations for the Green functions, find
the energies of the bound states, and calculate the supercurrent
through such junction. In Section \ref{sec-IV} we calculate the
current-voltage curve for a contact between the semi-infinite
normal and the proximity-induced superconducting regions. Our
results are summarized in Section \ref{sec-disc}.

\section{Model}

\begin{figure}[t]
\centering
\includegraphics[width=0.5\linewidth]{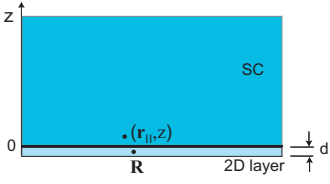}
\caption{Superconducting correlations in the 2D system under the
superconductor are induced through a thin
insulator.}\label{fig-model}
\end{figure}

Consider a 2D electron layer placed under a bulk superconductor
and coupled to it via tunnelling through a thin insulator coating,
Fig.~\ref{fig-model}. The Hamiltonian of the system has the form
\begin{equation}
\hat H = \hat H_S + \hat H_{2D} +\hat H_{T}\ .
\end{equation}
In the superconductor
\begin{eqnarray}
\hat H_S= \int d^3 r \left[ \sum_\alpha \hat \Psi^\dagger_\alpha
({\bf r})[\hat \epsilon_S -E_F] \hat \Psi_\alpha ({\bf
r})\right. \nonumber \\
+\left. \Delta \hat \Psi^\dagger_\uparrow ({\bf r}) \hat
\Psi^\dagger_\downarrow ({\bf r}) + \Delta^* \hat \Psi_\downarrow
({\bf r}) \hat \Psi_\uparrow ({\bf r})\right]\ ,
\end{eqnarray}
where $E_F$ is the chemical potential in the superconductor, $\hat
\epsilon_s $ is the kinetic energy operator. For parabolic
spectrum it is $ \hat \epsilon_S =\frac{1}{2m}\left(-i\hbar
\nabla-\frac{e}{c}{\bf A}\right)^2 $. The coordinate ${\bf r}$ is
a 3D vector which belongs to the superconductor. In the 2D layer
\begin{equation}
\hat H_{2D}= d\int d^2 R\,  \sum_\alpha \hat a^\dagger_\alpha
({\bf R}) \left[\hat \epsilon_{2D}({\bf R})-E_F\right] \hat
a_\alpha ({\bf R})\ ,
\end{equation}
where $d$ is the layer thickness, $\hat \epsilon _{2D}({\bf R})$
is the kinetic energy operator in the 2D layer. The coordinate
${\bf R}$ is a 2D vector which belongs to the layer. The creation
and annihilation operators in the layer $\hat a^\dagger_\alpha
({\bf R})$, $\hat a_\alpha ({\bf R})$ are normalized to the layer
volume, such that the anti-commutator $ \left[\hat a_\alpha ({\bf
R}), \hat a^\dagger_\alpha ({\bf R}^\prime)\right]_+ =d^{-1}
\delta({\bf R} -{\bf R}^\prime) $. The tunnel Hamiltonian has the
form
\begin{eqnarray}
\hat H_T = d \sum _\alpha \int \left[ \hat\Psi_\alpha ^\dagger
({\bf r})T({\bf r},{\bf R})\hat a_\alpha ({\bf R})\right. \nonumber \\
+\left. \hat a_\alpha ^\dagger ({\bf R})T^\dagger ({\bf R},{\bf
r})\hat \Psi_\alpha ({\bf r})\right]\, d^3r \, d^2 R\ .
\end{eqnarray}
Here the coordinate ${\bf R}$ refers to the layer while coordinate
${\bf r}$ refers to the superconductor; the matrix element $T({\bf
r},{\bf R})$ describes tunnelling between the layer and the
superconductor, $ T^*({\bf r}, {\bf R})=T^\dagger ({\bf R},{\bf
r}) $. This model is similar to the model used in Refs.
\cite{Shiba,ArseevVolkov91} for an impurity in superconductor.

The Matsubara Green functions are
\begin{eqnarray*}
\left< T_\tau \hat a_\alpha({\bf R}_1) \hat a_\beta ^\dagger ({\bf
R}_2)\right>=
\delta_{\alpha\beta}G_{2D}({\bf R}_1,{\bf R}_2) \ ,  \\
\left< T_\tau \hat \Psi_\alpha({\bf r}_1) \hat a_\beta ^\dagger
({\bf R}_2)\right>=
\delta_{\alpha\beta}G_T({\bf r}_1,{\bf R}_2) \ ,  \\
\left< T_\tau \hat \Psi_\alpha({\bf r}_1) \hat \Psi_\beta ^\dagger
({\bf r}_2)\right>= \delta_{\alpha\beta}G_S({\bf r}_1,{\bf r}_2)\
,
\end{eqnarray*}
and
\begin{eqnarray*}
\left< T_\tau \hat a_\alpha({\bf R}_1) \hat a_\beta ({\bf
R}_2)\right>=i\sigma^{(y)}_{\alpha\beta}F_{2D}({\bf
R}_1,{\bf R}_2) \ , \\
\left< T_\tau \hat \Psi_\alpha({\bf r}_1) \hat a_\beta ({\bf
R}_2)\right>=i\sigma^{(y)}_{\alpha\beta}F_T({\bf
r}_1,{\bf R}_2) \ , \\
\left< T_\tau \hat \Psi_\alpha({\bf r}_1) \hat \Psi_\beta ({\bf
r}_2)\right>=i\sigma^{(y)}_{\alpha\beta}F_S({\bf r}_1,{\bf r}_2)\
,
\end{eqnarray*}
etc.
We introduce the Nambu matrixes
\[
\check H_S({\bf r}) =\left(\begin{array}{cc} \hat \epsilon_S -E_F & -\Delta \\
\Delta ^* &\hat \epsilon_S -E_F
\end{array}\right) \; , \; \check G =\left(\begin{array}{cc}
G&F\\-F^\dagger &\bar G\end{array}\right)\ ,
\]
and denote
\[
\check G_S^{-1}({\bf r}) =\check \tau _3 \hbar \frac{\partial
}{\partial \tau} +\check H_S({\bf r})
\]
the inverse operator in the superconductor. If needed, it can also
include the impurity self energy. Neglecting the back-action of
the thin 2D layer on the superconductor, we have for the
superconducting Green function
\[
\check G_S^{-1}({\bf r}_1)\check G_S({\bf r}_1,{\bf r}_2) =\check
1 \hbar\delta({\bf r}_1-{\bf r}_2)\delta(\tau_1-\tau_2)\ .
\]

Equations for the mixed Green functions $\check G_T({\bf r}_1,{\bf
R}_2) $ can be written in the form
\[
\check G_S^{-1}({\bf r}_1)\check G_T({\bf r}_1,{\bf R}_2)+d\!\!
\int \check T({\bf r}_1,{\bf R}^\prime) \check G_{2D}({\bf
R}^\prime ,{\bf R}_2)\, d^2 R^\prime =0
\]
where
\[
\check T({\bf r},{\bf R})= \left(\begin{array}{lr} T({\bf
r},{\bf R}) & 0 \\
0 & T^*({\bf r},{\bf R})\end{array}\right)\ .
\]
This gives
\begin{eqnarray}
&&\check G_T({\bf r}_1,{\bf R}_2)\nonumber \\
&&=-d\!\! \int \check G_s({\bf r}_1,{\bf r}^\prime) \check T({\bf
r}^\prime ,{\bf R}^\prime) \check G_{2D}({\bf R}^\prime ,{\bf
R}_2)\, d^2 R^\prime\, d^3 r^\prime \ . \qquad \label{G-across}
\end{eqnarray}

Equations for the Green functions in the layer can be written as
\begin{eqnarray*}
\check G_{2D}^{-1}({\bf R}_1)\check G_{2D}({\bf R}_1,{\bf R}_2)
+\!\! \int \check T^\dagger ({\bf R}_1,{\bf r}^\prime) \check
G_T({\bf
r}^\prime ,{\bf R}_2)\, d^3 r^\prime \\
 =\check 1 d^{-1} \hbar\delta
({\bf R}_1 -{\bf R}_2)\delta(\tau_1-\tau_2)
\end{eqnarray*}
where the inverse operator in the layer is introduced,
\begin{equation}
\check G_{2D}^{-1}({\bf R})=\hbar \check\tau_3\frac{\partial
}{\partial \tau } + [\check \epsilon_{2D}({\bf R})-E_F]\ .
\end{equation}
Here the kinetic energy operator is
\[
\check
\epsilon_{2D}=\frac{1}{2m}\left(-i\hbar\frac{\partial}{\partial
{\bf R}} -\check \tau_3 \frac{e}{c}{\bf A}\right)^2 +\epsilon_0\ ,
\]
while $\epsilon_0$ is the shift of the bottom of the 2D conduction
band measured from that in the superconductor.

Using Eq. (\ref{G-across}) this equation becomes
\begin{eqnarray}
&&\check G_{2D}^{-1}({\bf R}_1)\check G_{2D}({\bf R}_1,{\bf R}_2)
-\int \check \Sigma _T({\bf R}_1,{\bf R}^\prime ) \nonumber
\\
&&\times \check G_{2D}({\bf R}^\prime ,{\bf R}_2)\, d^2 R^\prime
=\check 1 \hbar d^{-1} \delta ({\bf R}_1 -{\bf
R}_2)\delta(\tau_1-\tau_2)\qquad \label{eqG-coord}
\end{eqnarray}
where
\[
\check \Sigma _T({\bf R}_1,{\bf R}^\prime )=d\!\! \int \check
T^\dagger ({\bf R}_1,{\bf r}^\prime) \check G_S({\bf r}^\prime
,{\bf r}^{\prime \prime})\check T({\bf r}^{\prime \prime} ,{\bf
R}^\prime )\, d^3 r^\prime \, d^3 r^{\prime \prime}.
\]

We assume a site-to-site tunnelling which does not conserve
momentum, $ T({\bf r},{\bf R})=t \delta({\bf r}_\parallel -{\bf
R})\delta (z) $, etc., where $t$ is real and does not depend on
${\bf R}$. Here $\delta(z)$ selects an average value of a function
at a distance of the order of inter-atomic scale near the surface.
As a result,
\begin{equation}
\!\!\!\! \check \Sigma _T({\bf R}_i,{\bf R}_j)=  d t^2 \!\!
\left(\begin{array}{cc} G_S({\bf R}_i,{\bf R}_j; 0)
\ &  F_S({\bf R}_i,{\bf R}_j; 0) \\
- F^\dagger _S({\bf R}_i,{\bf R}_j;0) \ &  \bar G_S({\bf R}_i,{\bf
R}_j;0)
\end{array}\right) \label{Sigma-gen}
\end{equation}
where $({\bf R}_i; 0)$ means that ${\bf r}_\parallel = {\bf R}_i$
and $z_i=z_j=0$.

The coordinate dependence of the Green functions has the form
\cite{GorkovKopnin73}
\begin{eqnarray}
\check G_S({\bf r}_i,{\bf r}_j)= \frac{m e^{-|{\bf r}_i-{\bf
r}_j|/\ell }}{2\pi\hbar^2 |{\bf r}_i-{\bf r}_j|}\left[\check 1
\cos(p_F |{\bf r}_i-{\bf
r}_j|/\hbar )\right. \nonumber \\
+\left. i \sin (p_F|{\bf r}_i-{\bf r}_j|/\hbar) \check g_S(\hat
{\bf n}, {\bf r}) \right]\ . \label{G-coord}
\end{eqnarray}
Here $ \ell$ is the mean free path is the superconductor, $\hat
{\bf n} =({\bf r}_i-{\bf r}_j)/|({\bf r}_i-{\bf r}_j)|$ is a unit
vector, ${\bf r}=({\bf r}_i+{\bf r}_j)/2$,  and $\check g_S$ is
the quasiclassical superconducting Green function integrated over
the energy variable, Eq. (\ref{quasiclassG}),
\begin{equation}
\check g_S({\bf p}_F,{\bf r})=\int \frac{d\epsilon_s}{i\pi \hbar}
\check G_S({\bf p},{\bf r}) \label{quasiclassG}
\end{equation}
which depends only on the direction of the momentum ${\bf
p}_F=p_F{\bf n}$.

The self-energy $\check \Sigma _T({\bf R}_i,{\bf R}_l)$ in Eq.
(\ref{Sigma-gen}) oscillates as $\exp(ip_F|{\bf R}_j-{\bf
R}_l|/\hbar)$ while the Green function contains $\exp[i {\bf
p}({\bf R}_l-{\bf R}_j^\prime)/\hbar]=\exp[i {\bf p}({\bf
R}_l-{\bf R}_j)/\hbar+i {\bf p}({\bf R}_j-{\bf
R}_j^\prime)/\hbar]$ where $|{\bf p}|$ is close to the Fermi
momentum $p_{2D}$ in the 2D layer. If $p_{2D}\ne p_F$, the
integral in Eq. (\ref{eqG-coord}) converges at $R=|{\bf R}_j-{\bf
R}_l|\sim a$ where $a$ is the interatomic distance. Therefore, one
can invoke the approximation
\[
\Sigma _T ({\bf R}_j,{\bf R}_{l} )= \Sigma ({\bf R}_j)\delta({\bf
R}_j-{\bf R}_l)\ .
\]
The first term in Eq. (\ref{G-coord}) diverges for $|{\bf R}-{\bf
R}^\prime|\to 0$, but it does not depend on the superconducting
state. It is real and has the form of an effective potential. We
can include it into the shift of the bottom of the conduction band
$\epsilon_0$ or into the chemical potential assuming that $\mu$
already accounts for it. The remaining term for $|{\bf R}-{\bf
R}^\prime|\to 0$ gives
\begin{equation}
\check \Sigma({\bf R}) =i\Gamma \left<\check g_S({\bf R};0)\right>
\ .
\end{equation}
Here we introduce the tunnelling rate $\Gamma =\pi \nu_3d s_0 t^2$
where $\nu_3=mp_F/2\pi^2\hbar^3 $ is the 3D density of states,
$s_0\sim a^2$ is of the order of the area of the unit cell in the
layer; the overall order of magnitude is $\Gamma \sim t^2/E_F$.
Angular brackets denote averaging over momentum directions. The
quasiclassical Green function $\check g_S({\bf R};0)$ taken at the
superconductor/2D-layer interface is the only characteristic which
is needed in our model to account of the properties of the bulk
superconductor. Since the quasiclassical function varies over
distances of the order of superconducting coherence length, the
exact atomic-scale boundary conditions at the interface for the
microscopic wave functions are not critical for our model.

We write the final equation in 2D layer for the real-time Green
functions making the analytical continuation of the Matsubara
functions onto the real-frequency axis. We introduce the Keldysh
matrixes
\[
{\cal G}=\left(\begin{array}{cc} \check G^R & \check G^K \\
0 & \check G^A \end{array}\right)\ , \; {\cal S}=
\left(\begin{array}{cc} \check \Sigma^R & \check \Sigma^K \\
0 & \check \Sigma^A \end{array}\right)\ .
\]
The  equations become (the index $2D$ is dropped)
\begin{eqnarray}
\left(\hat \epsilon_{2D} -E_F -\check \tau_3 \epsilon \right)
{\cal G}_{\epsilon,\epsilon^\prime} ({\bf R},{\bf R}^\prime)
-\left [{\cal S}({\bf R}) {\cal G}({\bf R},{\bf
R}^\prime)\right]_{\epsilon,\epsilon^\prime}\nonumber \\
= \check 1 \hbar^2 d^{-1}\delta({\bf R}-{\bf R}^\prime)2\pi
\delta(\epsilon -\epsilon^\prime) \; \label{GreenEquation}
\end{eqnarray}
where
\[
\left [{\cal S}({\bf R}) {\cal G}({\bf R},{\bf
R}^\prime)\right]_{\epsilon,\epsilon^\prime}=\int {\cal
S}_{\epsilon,\epsilon_1} ({\bf R}) {\cal
G}_{\epsilon_1,\epsilon^\prime}({\bf R},{\bf
R}^\prime)\frac{d\epsilon_1}{2\pi\hbar}\ .
\]

\section{S/2D/S structure} \label{sec-SDS}

\subsection{Green functions}

\begin{figure}[b]
\centering
\includegraphics[width=1.00\linewidth]{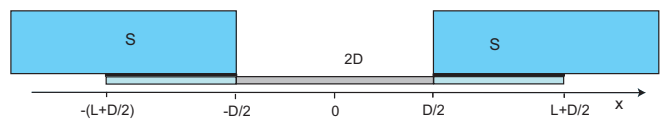}
\caption{Contact made of two superconducting electrodes placed on
top of the 2D electron gas at a distance $D$ from each other. Each
electrode covers a length $L$ of the 2D
layer.}\label{fig-junctionSNS}
\end{figure}

In this paper we restrict ourselves to stationary problems and put
$ \check G^{R(A)}_{\epsilon,\epsilon^\prime}= \check
G^{R(A)}_{\epsilon}2\pi \hbar  \delta(\epsilon-\epsilon^\prime) $.
In the following Sections we apply our model to the S/2D/S
structures which are made of a ballistic 2D electron gas placed in
a contact with bulk superconducting leads (S). One of the examples
is shown in Fig.~\ref{fig-junctionSNS}. A generic structure of
this type consists of a layer of 2D electron gas which is infinite
in $y$ direction and has boundaries at certain $x$ where ${\bf
R}=(x, y)$. Transforming to the plane waves along the $y$
direction,
\[
\check G^R_{\epsilon} ({\bf R},{\bf R}^\prime)=\int
e^{ip_y(y-y^\prime)/\hbar}\check G^R_{\epsilon,p_y}
(x,x^\prime)\frac{dp_y}{2\pi\hbar}\ ,
\]
we observe that, as functions of $x$ for $x\ne x^\prime$, the
retarded Green functions $G^R_{\epsilon,p_y} (x,x^\prime)$ and
$F^{\dagger R}_{\epsilon,p_y} (x,x^\prime)$ satisfy the set of 2
linear homogeneous second-order differential equations which
follow from Eq. (\ref{GreenEquation}),
\begin{eqnarray}
\left[-\frac{\hbar^2}{2m}\frac{d^2}{dx^2}-\mu_x\right] u(x)
-(\epsilon+\eta_1)u(x)+\eta_2 v(x) =0  ,\quad \label{eqxuv1} \\
\left[\frac{\hbar^2}{2m}\frac{d^2}{dx^2}+\mu_x \right]v(x)
-(\epsilon+\eta_1 )v(x) +\eta_2^\dagger u(x) =0  ,\quad
\label{eqxuv4}
\end{eqnarray}
where $\mu=E_F-\epsilon_0$ is the chemical potential in the 2D gas
measured from the bottom of the conduction band, $ \mu_x =\mu -
p_y^2/2m$, and
\begin{equation}
\eta_1=i\Gamma \left< g^R_\epsilon\right>\ , \; \eta_2=i\Gamma
\left< f^R_\epsilon\right>\ , \; \eta_2^\dagger =i\Gamma \left<
f^{\dagger R}_\epsilon\right>\ .
\end{equation}
Eqs. (\ref{eqxuv1}), (\ref{eqxuv4}) have 4 independent solutions,
\[
\check \Psi^{(i)}(x) =\left( \begin{array}{c} u^{(i)}(x) \\
v^{(i)}(x)
\end{array}\right)\ , \; i=1,2,3,4\ .
\]
Using the standard method in the theory of differential equations,
one can write the Green functions as sums
\begin{eqnarray*}
\left( \begin{array}{c}  G(x,x^\prime) \\ F^{\dagger} (x,x^\prime)
\end{array}\right) = C_1(x^\prime)\check \Psi^{(1)}(x) + C_2(x^\prime)
\check \Psi^{(2)}(x) ,\; x>x^\prime , \\
\left( \begin{array}{c} G(x,x^\prime) \\ F^{\dagger} (x,x^\prime)
\end{array}\right) = -C_3(x^\prime)\check \Psi^{(3}(x) - C_4(x^\prime)
\check \Psi^{(4)}(x) , \; x<x^\prime  .
\end{eqnarray*}
The functions $\check \Psi^{(i)}(x)$ are chosen such that the
retarded Green functions are regular in the upper half-plane,
${\rm Im}\, \epsilon >0$. The coefficients $C_k(x^\prime)$ do not
depend on $x$ but can be functions of $x^\prime$. We find from Eq.
(\ref{GreenEquation})
\begin{eqnarray*}
G(x+0,x)-G(x-0,x)=0\ , \\
\frac{d}{dx}G(x+0,x)-\frac{d}{dx}G(x-0,x) =-\frac{2m}{\hbar d}\ , \\
F^\dagger (x+0,x)-F^\dagger (x-0,x)=0\ , \\
\frac{d}{dx}F^\dagger(x+0,x)-\frac{d}{dx}F^\dagger(x-0,x)=0\ .
\end{eqnarray*}
Let us define the  vectors
\[
C_i= \left(\begin{array}{c}C_1\\C_2\\C_3\\C_4\end{array}\right)\ ,
\; R_k= -\frac{2m}{\hbar d}\left(\begin{array}{c}0\\1
\\0\\0\end{array}\right)
\]
and the matrix $U^i_k$ where
\[
U^i_1=u^{(i)}\ ,\; U^{i}_2= du^{(i)}/dx\ , U_3^{i}=v^{(i)}\ ,\;
U^{i}_4= dv^{(i)}/dx\ .
\]
The upper index numerates columns, while the lower numerates rows.
Equations take the form $ \sum_{i} U^{i}_k C_i=R_k $. The solution
is
\begin{equation}
C_i(x) = W^{-1} \sum_k A^k_{i}(x)R_k \ . \label{Green-C}
\end{equation}
Here $ W ={\rm det}\, [U^{i}_k] $ is the Wronskian and $ A^i_k(x)
= (-1)^{i+k}W^i_k(x) $ where $W_k^i(x)$ is a minor of ${\rm det}\,
[U^{i}_k]$. The Wronskian is independent of coordinates.

\subsection{Reflection coefficients} \label{subsec-reflection}

Consider structure shown in Fig. \ref{fig-junctionSNS}. The right
superconductor has the phase $\phi/2$ while the left one has
$-\phi/2$. The four independent functions needed for the Green
functions can be constructed as follows. The first two wave
functions inside the 2D layer $-D/2 <x<D/2$ contain waves incident
from the left and reflected from the right boundary,
\begin{eqnarray}
\check \Psi^{(1)}(x)&=&e^{ik_+ (x-\frac{D}{2})}\check \Psi_p +
r_A^R e^{ik_- (x-\frac{D}{2})}\check \Psi_h \nonumber \\
&& + r_N^R e^{-ik_+ (x-\frac{D}{2})}\check \Psi_p\ ,
\label{function-1}\\
\check \Psi^{(2)} (x)&=& e^{-ik_- (x-\frac{D}{2})}\check \Psi_h +
\bar r_A^R e^{-ik_+ (x-\frac{D}{2})}\check \Psi_p \nonumber \\
&& + \bar r_N^R e^{ik_- (x-\frac{D}{2})}\check \Psi_h\ .
\label{function-2}
\end{eqnarray}
Here $ k_\pm = k_{x}\pm \epsilon /\hbar v_x $, while $k_x$ and
$v_x$ are the $x$ components of the Fermi wave vector and Fermi
velocity $\hbar k_x=mv_x$ in the 2D gas, $\hbar^2 k_x^2/2m
=\mu_x$; the particle and hole Nambu vectors are
\[
\check \Psi_p =\left(\begin{array}{c} 1\\0\end{array}\right)\ , \;
\check \Psi_h =\left(\begin{array}{c} 0\\1\end{array}\right)\ .
\]
The set of two other functions is obtained using reflection from
the left boundary,
\begin{eqnarray}
\check \Psi^{(3)}(x)&=&e^{-ik_+ (x+\frac{D}{2})}\check \Psi_p +
r_A^L e^{-ik_- (x+\frac{D}{2})}\check \Psi_h \nonumber \\
&&+ r_N^L e^{ik_+
(x+\frac{D}{2})}\check \Psi_p \ , \label{function-3}\\
\check \Psi^{(4)}(x)&=&e^{ik_- (x+\frac{D}{2})}\check \Psi_h +
\bar r_A^L e^{ik_+ (x+\frac{D}{2})}\check \Psi_p \nonumber \\
&& + \bar r_N^L e^{-ik_- (x+\frac{D}{2})}\check \Psi_h\ .
\label{function-4}
\end{eqnarray}
In the right induced-superconductivity region $D/2<x<L+D/2$,  the
four functions satisfying Eqs. (\ref{eqxuv1}) - (\ref{eqxuv4}) are
linear combinations of
\begin{eqnarray*}
e^{ik_+^s (x-L-\frac{D}{2})}\left(\begin{array}{c} u
e^{i\frac{\phi}{4}}
\\ v e^{-i\frac{\phi}{4}}  \end{array}\right) , \; e^{-ik_+^s (x-L-\frac{D}{2})}
\left(\begin{array}{c}
u e^{i\frac{\phi}{4}} \\ v e^{-i\frac{\phi}{4}}  \end{array}\right), \nonumber \\
e^{ik_-^s (x-L-\frac{D}{2})}\left(\begin{array}{c} v e^{i\frac{\phi}{4}} \\
u e^{-i\frac{\phi}{4}} \end{array}\right)  , \; e^{-ik_-^s
(x-L-\frac{D}{2})}\left(\begin{array}{c} v e^{i\frac{\phi}{4}} \\
u e^{-i\frac{\phi}{4}}
\end{array}\right).
\end{eqnarray*}
The functions in the left superconducting region are obtained by
replacing $\phi$ with $-\phi$ and $(L+D/2)$ with $-(L+D/2)$. Note
that the momentum in the induced-superconductivity region,
\begin{equation}
k^s_\pm =k_{x}\pm \frac{1}{\hbar v_x}\sqrt{
(\epsilon+\eta_1)^2-\eta_2\eta_2^\dagger}\ ,  \label{pspm}
\end{equation}
has the same Fermi-momentum projection on the $x$ coordinate as in
the normal region. Thus the normal reflection at $x=\pm D/2$ does
not occur. The coherence factors are
\begin{equation}
u(\epsilon)=\frac{1}{\sqrt{2}}\sqrt{
1+\frac{\xi}{\epsilon+\eta_1}}\ , \;
v(\epsilon)=\frac{1}{\sqrt{2}}\sqrt{
1-\frac{\xi}{\epsilon+\eta_1}} \ . \label{exp-v}
\end{equation}
Here
\begin{equation}
\xi^2= (\epsilon+\eta_1)^2-\eta_2\eta_2^\dagger =\tilde
\epsilon^2-\Gamma^2\ . \label{exp-xi}
\end{equation}
since $\eta_1^2-\eta_2\eta_2^\dagger =-\Gamma^2$. Moreover
\[
g^R=\frac{\epsilon}{\sqrt{\epsilon^2-\Delta^2}} , \;
f^R=\frac{\Delta}{\sqrt{\epsilon^2-\Delta^2}}, \; \tilde
\epsilon^2 =\epsilon^2+\frac{2i\Gamma\epsilon^2}{\sqrt{\epsilon^2
-|\Delta|^2}}.
\]
The square roots are defined as analytic functions in the plane of
complex $\epsilon$ with cuts from $|\Delta|$ to $\infty$ and from
$-\infty$ to $-|\Delta|$ taken at the upper bank of the cut for
$\epsilon >|\Delta|$. The usual BCS formulas are recovered if one
uses $ \epsilon\to \epsilon+\eta_1\ , \; \Delta \to \eta_2\ , \;
\Delta^* \to \eta_2^\dagger $ such that the BCS gap $\Delta|^2$ is
replaced with $ \Gamma^2 |\Delta|^2/(|\Delta|^2-\epsilon^2) $.

The boundary conditions require that $ u^{(i)}=v^{(i)}=0$ at $
x=\pm (L+D/2)$ together with continuity of $ u^{(i)}\, , v^{(i)} \
, \; du^{(i)}/dx\ , \;  dv^{(i)}/dx$ at $ x=\pm D/2$. This yields
\begin{eqnarray}
r_N^R&=&r_N e^{i\delta}\ , \; \bar r_N^R=r_N e^{-i\delta}\ , \label{rN-R-2}\\
r_A^R&=&r_A e^{-i\phi/2}\ , \; \bar r_A^R= r_A e^{i\phi/2}\ ,
\label{rA-R-2} \\
r_N^L &=&r_Ne^{i\delta}\ ,  \bar r_N^L =r_Ne^{-i\delta} \ , \label{rN-L} \\
r_A^L &=&r_Ae^{i\phi/2}\ , \bar r_A^L =r_Ae^{-i\phi/2}\ ,
\label{rA-L}
\end{eqnarray}
where
\[
r_N \! =-\frac{(u^2-v^2)e^{i(\delta_+ -\delta_-)}}{u^2-v^2
e^{2i(\delta_+
-\delta_-)}} 
\ , \;
r_A \! =\frac{uv[1-e^{2i(\delta_+ -\delta_-)}]}{u^2-v^2
e^{2i(\delta_+ -\delta_-)}}\ . 
\]
Here we denote $\delta_\pm = k_\pm ^sL$ and $\delta=\delta_+
+\delta_-=2k_{x}L$, while
\begin{equation}
\delta_+ -\delta_-=(2L/\hbar v_{x})\sqrt{\tilde
\epsilon^2-\Gamma^2}\ .
\end{equation}
One can check that the Green functions defined according to Eq.
(\ref{Green-C}) with the basis functions Eqs.
(\ref{function-1})--(\ref{function-4}) are regular in the upper
half-plane of complex $\epsilon$. For $\epsilon <\Delta$ the
coefficients satisfy
\begin{equation}
|r_N|^2 +|r_A|^2 =1 \ , \; r_N^* r_A +r_A^* r_N =0 \label{norm1}
\end{equation}
because there is no quasiparticle flux into the superconductor.

As in Ref.~\cite{KM-etal-05} one can define an equivalent barrier
hight $Z$ associated with the end of the conduction layer,
\begin{equation}
Z/\sqrt{1+Z^2}=e^{i(\delta_+ -\delta_-)}\ ,  \label{Z}
\end{equation}
such that the coefficients have their standard form
\cite{BTK82,KMV2006}
\begin{eqnarray}
r_A\! =\frac{uv}{u^2\!+\!(u^2-v^2)Z^2} , \; r_N\! =\!
-\frac{(u^2\!-\!v^2)Z\sqrt{1 +Z^2}}{u^2\!+ (u^2-v^2)Z^2}.\quad
\label{rAN}
\end{eqnarray}
The limit $L=0$ corresponds to strong normal reflection, $Z\to
\infty$ and $|r_N|=1$, while $r_A=0$. With this definition of
effective barriers, the problems of the bound-state spectrum and
of the supercurrent in many respects reduce to the corresponding
problems in double-barrier superconductor/normal/superconductor
(SINIS) structures.

\begin{figure}[b]
\centering
\includegraphics[width=0.5\linewidth]{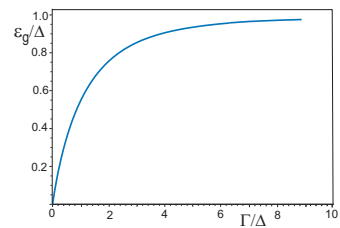}
\caption{Induced gap as a function of the tunneling
rate.}\label{fig-gap}
\end{figure}

Let us denote $\epsilon_g<\Delta$ the energy at which the square
root in Eq. (\ref{pspm}) vanishes, i.e., at which $\xi$ defined by
Eq. (\ref{exp-xi}) turns to zero,
\begin{equation}
\epsilon_g^2\left( 1+ 2\Gamma/\sqrt{\Delta^2-\epsilon_g^2}
\right)-\Gamma^2=0\ .
\end{equation}
The energy $\epsilon_g$ is the induced gap in the 2D layer. For
$\Gamma \ll \Delta$ we have $\epsilon_g =\Gamma$; if $\Gamma \gg
\Delta$ one has $\epsilon_g =\Delta(1-2\Delta^2/\Gamma^2)$. The
induced gap as a function of $\Gamma$ is shown in
Fig.~\ref{fig-gap}. For a 2D gas in a direct contact with a
superconductor, the induced gap was obtained in
Ref.\cite{AVolkov95}.

If $\epsilon^2 <\epsilon_g^2$, both the square root in Eq.
(\ref{pspm}) and $\xi$ are imaginary. The value $Z$ is real and
satisfies
\begin{equation}
Z/\sqrt{1+Z^2} =\exp\left[-2(L/\hbar v_x)\sqrt{\Gamma^2-\tilde
\epsilon^2}\right]\ .
\end{equation}
For large $L$ we have $Z=0$; the particles do not feel the dead
end of the superconducting region, such that the normal reflection
is absent, while $|r_A|=1$. This limit is realized when $L\gg
\xi_{2D}$ where
\[
\xi_{2D} = \hbar v_{2D}/\epsilon_g
\]
is the coherence length of the induced superconductivity, $v_{2D}$
is the Fermi velocity in the 2D system $v^2_{2D}/2m=\mu$. The
limit of long leads is rather hard to realize because $\xi_{2D}$
is considerably longer than $\xi_0$ in a bulk superconductor. On
the contrary, for short leads, $L=0$, there exists only normal
reflection, $Z\to \infty$, and $|r_N|=1$.

If $\epsilon_g<\epsilon <|\Delta|$ both the phases $\delta_\pm$
and the coherence factors $u$, $v$ are real, while $Z$ is complex.
The Andreev reflection vanishes while $|r_N|=1$ if $ \delta_+
-\delta_- =\pi n $.

For $\epsilon >|\Delta|$ the coefficients still have the form of
Eq. (\ref{rAN}) with a complex effective barrier strength $Z$. For
large $\epsilon\gg |\Delta|$ one has $\tilde \epsilon =\epsilon
+i\Gamma $. The effective barrier height saturates at $Z\ne 0$,
while $v\to 0$ and thus $r_A \to 0$. At the same time
\[
|r_N|^2=\exp \left[-4L\Gamma/\hbar v_x\right]\ .
\]
Equation (\ref{norm1}) no longer holds because the quasiparticles
escape into the superconductor.

\subsection{Bound states. Supercurrent}

Using the functions Eqs. (\ref{function-1}) -- (\ref{function-4})
and Eqs. (\ref{rN-R-2})--(\ref{rA-L}), we find the Wronskian
\begin{eqnarray*}
W&=&16k_+ k_- e^{2i\gamma} \left[ \sin^2(\beta+ \gamma ) -|r_A|^2
\cos^2 (\phi/2)\right. \\
&&\left. - |r_N|^2 \sin^2(\alpha+\delta)\right]
\end{eqnarray*}
where $ \alpha =k_{x}D$, $\beta =\epsilon D/\hbar v_{x} $, while
$\gamma$ is the phase of the reflection coefficient $r_N$, i.e., $
e^{2i\gamma}= r_N/r_N^{*} $. The spectrum is determined by $W=0$,
which gives
\begin{equation}
\sin^2(\beta+ \gamma ) =|r_A|^2 \cos^2 (\phi/2) + |r_N|^2
\sin^2\alpha^\prime  \label{spectrum}
\end{equation}
where $\alpha^\prime =\alpha+\delta$. This agrees with the
previous calculations of Ref.~\cite{KMV2006} for ballistic SINIS
contacts. Here we concentrate on short contacts such that $\beta
\ll 1 $. Neglecting $\beta$ and keeping $L$ finite implies that
one has to assume $L\gg D$.

If the energy is above the induced gap, $\epsilon_g<\epsilon
<\Delta$, the combination $\delta_+ -\delta_-$, as well as $u$ and
$v$ are real. For energies below the induced gap, $\epsilon
<\epsilon_g$, the barrier strength $Z$, Eq. (\ref{Z}), is real,
while $u=v^*$. Therefore, Eq. (\ref{spectrum}) yields for the
bound state energy
\begin{eqnarray}
\frac{(\epsilon+\eta_1)^2}{\eta_2\eta_2^\dagger}=1-
\frac{\sin^2(\phi/2)}{1+A}\label{E<}
\end{eqnarray}
where
\begin{eqnarray}
A&=&-\frac{\sin^2\alpha^\prime }{\sin^2[2L\sqrt{\tilde
\epsilon^2-\Gamma^2}/\hbar v_x]}\ , \; \epsilon_g<\epsilon <\Delta \label{A>}\\
A&=& \frac{\sin^2\alpha^\prime }{\sinh^2[2L\sqrt{\Gamma^2-\tilde
\epsilon^2}/\hbar v_x]}\ , \; \epsilon <\epsilon_g \label{A<}
\end{eqnarray}
Equation (\ref{A>}) goes into Eq. (\ref{A<}) if one continues
analytically the phase difference $\delta_+ -\delta_-$ as a
function of $\epsilon$ around the induced gap $\epsilon_g$.

For energies below $\Delta$ qusiparticles cannot escape into the
superconductors and form the bound states as a result of two
processes: the Andreev reflection at the boundary between the
normal region and the region with induced-superconductivity plus
the normal reflection at the ends of the 2D layer. The spectrum is
determined by Eq. (\ref{E<}); it consists of a series of levels
for each $p_y$ and $\phi$ spaced with a distance $\sim \hbar
v_x/L$ and filling the interval $0<\epsilon<\Delta$. The lowest
energy level lies below the induced gap if $(2L\Gamma/\hbar v_x)
\sin (\phi/2) > \sin \alpha^\prime $. For short leads,
$2L\Gamma/\hbar v_x \ll \sin \alpha^\prime $, the energies are
close to the levels of geometrical quantization in a potential
well of length $d+2L$, i.e., they satisfy $\delta_+-\delta_- \pm
\alpha^\prime =\pi n$. If $L\ll \hbar v_x/\Delta$, there is only
one bound state with energy $\epsilon <\Delta$.

The supercurrent is\cite{Beenakker91}
\begin{equation}
I=-\frac{\pi}{eR_0}\int_{-\pi/2}^{\pi/2} \cos\theta \,
d\theta\sum_{n} \frac{\partial \epsilon_n}{\partial \phi}
\tanh\frac{\epsilon_n}{2T}\ . \label{s-current}
\end{equation}
Here
\[
R_0^{-1}=2 e^2 w\nu_{2}v_{2D}/\pi =G_0wp_{2D}/\pi\hbar
\]
is the Sharvin conductance of the contact, $G_0=e^2/\pi \hbar$ is
the conductance quantum, $p_{2D}=mv_{2D}$ is the Fermi momentum in
the layer, and $ \nu_{2}= m/2\pi\hbar^2 $ is the normal density of
states in the 2D gas. The conductance is proportional to the
number of modes in the 2D channel of width $w$ in the $y$
direction. The sum collects all bound states with energies
$\epsilon<\Delta$ for given $p_y=p_{2D}\sin \theta$. The continuum
states with $\epsilon >\Delta$ give the contribution proportional
to the length $D$ and can thus be neglected.

\subsubsection{Short leads}

Consider the case of short leads, $L\Gamma /\hbar v_x \ll 1$ for
$\Gamma \ll \Delta$. This is the most practical situation in the
experimental devices. If $\hbar v_x/\Delta \ll L \ll \hbar
v_x/\Gamma$, there is a series of bound states with energies
$0<\epsilon_n<\Delta$. As discussed above, they are close to the
levels of dimensional quantization. Putting $x=(2L\Gamma/\hbar
v_x)\sqrt{\epsilon^2/\Gamma^2-1}$ we observe that $x=x_n+\delta
x_n$ where
\begin{equation}
\sin^2 x_n =\sin^2\alpha^\prime  \ , \label{dimens}
\end{equation}
and
\begin{equation}
\delta x_n=-\left(\frac{2L\Gamma}{\hbar v_x}\right)^2
\sin^2(\phi/2) \frac{\tan x_n}{2x_n^2} \label{E-off2}
\end{equation}
which follows from Eqs. (\ref{E<}), (\ref{A>}). Equation
(\ref{dimens}) gives
\begin{equation}
x_n =  \left\{ \begin{array}{cl} \pm x_0 +\pi n ,&  n=1,2, \ldots \ , \\
 x_0 , &  n=0\ . \end{array}\right. \label{E-off1}
\end{equation}
Here $x_0= \arcsin|\sin \alpha^\prime |$ is the smallest root,
$0<x_0<\pi /2$, of Eq. (\ref{dimens}). This solution holds as long
as $\delta x_n \ll x_n$, i.e., for all states with $n\ne 0$. It
also holds for the lowest energy state, as long as the level is
not very close to the resonance, $|\sin\alpha^\prime |\gg
(2L\Gamma/\hbar v_x) $.

\begin{figure}[t]
\centering
\includegraphics[width=0.7\linewidth]{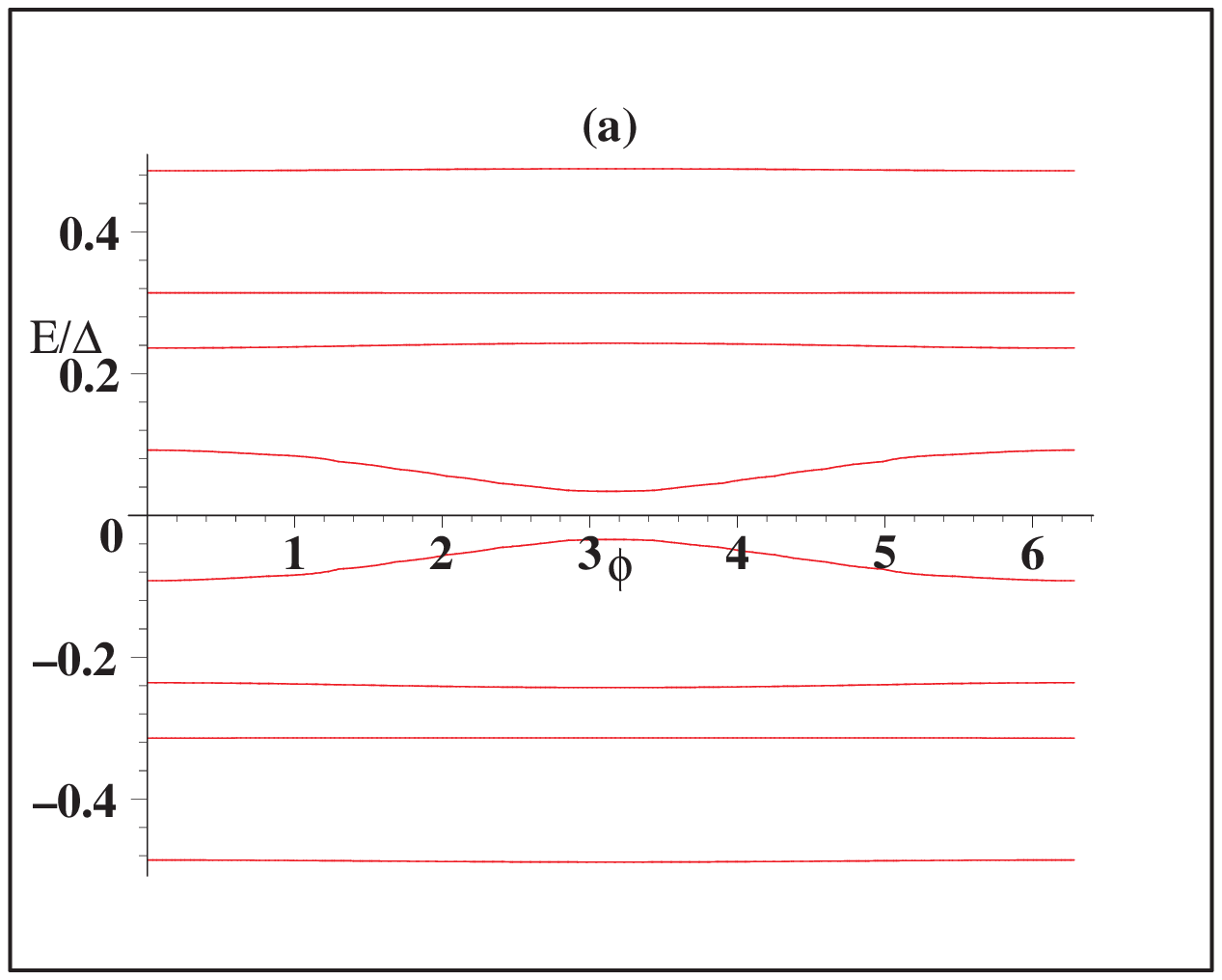}
\includegraphics[width=0.7\linewidth]{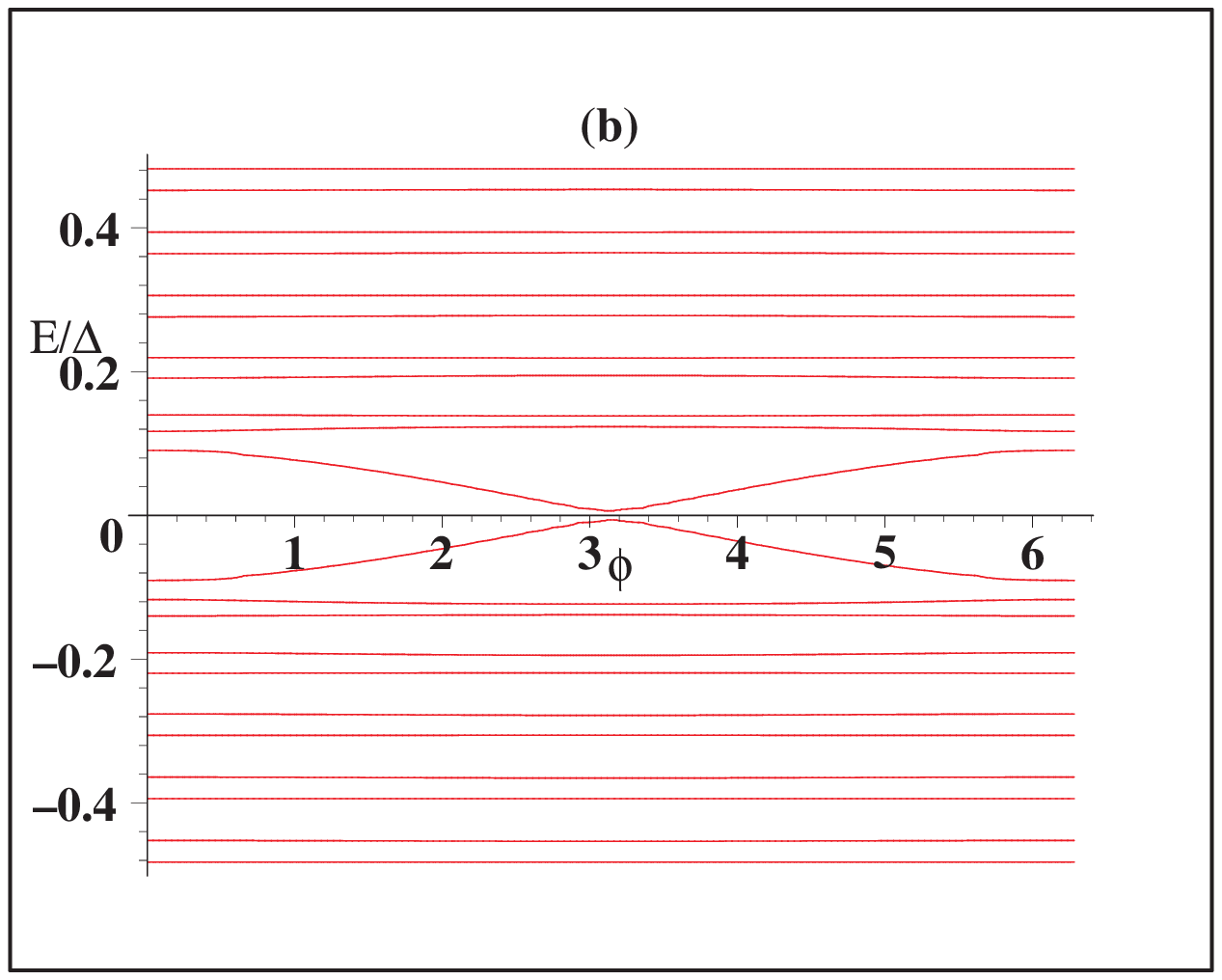}
\caption{Spectrum of bound states in a Josephson junction as a
function of superconducting phase difference. Here we put (a)
$\Gamma/\Delta=0.1$, $2L\Delta/\hbar v_F=10$, $\hbar k_F
v_F/\Delta=10$, $D\Delta/\hbar v_F=1$, $\theta=0$; (b)
$\Gamma/\Delta=0.1$, $2L\Delta/\hbar v_F=30$, $\hbar k_F
v_F/\Delta=10$, $D\Delta/\hbar v_F=1$,
$\theta=0$}\label{fig-boundstates}
\end{figure}

Close to the resonance, $|\sin\alpha^\prime |\ll 1$, the lowest
level can also be obtained by expanding the sine functions in Eqs.
(\ref{A>}), (\ref{A<}). Putting $\tilde \epsilon \approx \epsilon$
we find
\begin{eqnarray}
\epsilon^2 &=&\frac{\Delta^2+\Gamma^2(1+a^2)}{2}\nonumber \\
&&-\sqrt{\frac{[\Delta^2-\Gamma^2(1+a^2)]^2}{4}+\Delta^2 \Gamma^2
\sin^2(\phi/2)}\ ,\quad  \label{E-bound}
\end{eqnarray}
where $ a^2= \hbar^2 v_x^2\sin^2\alpha^\prime /(2L\Gamma)^2 $.
Close to the resonance, $a\Gamma \ll\Delta$,
\begin{equation}
\epsilon = \Gamma\sqrt{a^2+\cos^2(\phi/2)}\ . \label{E-res}
\end{equation}
This goes over into Eqs. (\ref{E-off1}), (\ref{E-off2}) when $a\gg
1$. The typical spectrum is shown in
Fig.~\ref{fig-boundstates}(a).

If the leads are very short, $L\ll \hbar v_x/\Delta$ then there
exists only one level. Except for a very narrow vicinity of
resonance, one has $a^2\Gamma^2 \gg \Delta^2$, thus the level is
close to the superconducting gap $\Delta$,
\begin{equation}
\epsilon^2= \Delta^2-(\Delta^2/a^2) \sin^2(\phi/2)\ .
\label{E-off3}
\end{equation}


To calculate the supercurrent Eq. (\ref{s-current}) we integrate
the contribution from each level over the incident angle $\theta$
defined according to $v_x=v_{2D}\cos \theta$. Since the resonance
form of the level, Eq. (\ref{E-res}), holds only in a very narrow
region of angles, its contribution to the current is small. The
current is thus mostly determined by the off-resonance levels,
Eqs. (\ref{E-off1}), (\ref{E-off2}), or (\ref{E-off3}) with
$\epsilon_n \gg \Gamma$. For $\hbar v_x/\Delta \ll L \ll \hbar
v_x/\Gamma $, there is a series of levels defined by Eqs.
(\ref{E-off1}), (\ref{E-off2}). The current is
\begin{eqnarray}
I\!\!&=&\!\!\frac{\pi}{eR_0}\left(\frac{L\Gamma^2}{\hbar
v_{2D}}\right)\sin \phi
\int_{0}^{\pi/2} \tan x_0\nonumber \\
&&\times \left[ F(x_0)+\sum_{n=1}^\infty [ F(\pi n +x_0)-F(\pi
n-x_0)]\right]\, d\theta \qquad \label{curr-int}
\end{eqnarray}
where
\[
F(x)= \frac{1}{x^2} \tanh\frac{\hbar v_x x}{4LT}\ .
\]
Due to a large argument in $\sin
(\alpha^\prime)=\sin[k_{2D}(d+2L)\cos \theta]$, the level $x_0$
oscillates rapidly assuming values within the interval
$0<x_0<\pi/2$ many times as the incident angle $\theta$ varies
from $0$ to $\pi/2$. Let us consider a function $f(x_0,\theta)$ of
the rapidly oscillating variable $x_0(\theta)$ and a slow variable
$\theta$ and define the average function $ \left<
f(\theta)\right>\delta \theta =\int_{\theta}^{\theta
+\delta\theta} f(x_0,\theta)\, d\theta $ where the integral is
taken over the full variation range of $x_0$ putting $dx_0 =
(d+2L)k_{2D}\sin \theta d\theta$. Since the range $0<x_0<\pi/2$
corresponds to a small variation $\delta \theta \ll 1$, one can
keep the slow variable $\theta$ constant during integration over
$dx_0$,
\[
\left< f(\theta)\right> =\frac{2}{\pi}\int_{0}^{\pi/2}
f(x_0,\theta)\, dx_0\ .
\]
The integral of the rapid function $f(x_0,\theta)$ can now be
replaced with the integral of the average function $\int_0^{\pi/2}
f(x_0,\theta)\, d\theta= \int_0^{\pi/2} \left< f(\theta)\right>\,
d\theta$.

For low temperatures, $T\ll \hbar v_{2D}/L$ one can simplify the
expression (\ref{curr-int}) for the current. We note that for $n
\ne 0$ the integral in $\left< \tan x_0 F(\pi n\pm x_0 )\right>$
diverges logarithmically at $x_0 \to \pi /2$; it should be cut off
when $\delta x_n \sim 1$, i.e., at $\pi/2 -x_0\sim
\left(L\Gamma\sin(\phi/2)/\hbar v_{x}\right)^2$. Therefore, within
the logarithmic accuracy,
\begin{eqnarray*}
\left< \tan x_0 F(\pi n\pm x_0 )\right>\!\! &=&\!\!
\left(\frac{4}{\pi}\right)^3 \ln
\left(\frac{\hbar v_{2D}}{L\Gamma}\right) \\
&&\times \frac{1 }{(2n\pm 1)^2}\tanh\frac{\pi\hbar v_{x} (2 n\pm
1)}{8LT}\ .
\end{eqnarray*}
For $n = 0$ the integral diverges logarithmically at $x_0 \to \pi
/2$. At the lower limit, $x_0\to 0$, the function $F(x)\tan x$
under the integral should be replaced with that containing the
resonance level taken from Eq. (\ref{E-res}). Therefore, we find
\[
\left< \tan x_0 F(x_0 )\right> =\left(\frac{4}{\pi}\right)^3 \ln
\left(\frac{\hbar v_{2D}}{L\Gamma}\right) \tanh\frac{\pi\hbar
v_x}{8LT} +\frac{2}{\pi}B(\theta)
\]
where
\begin{eqnarray*}
B (\theta)\!=\! \int _{x_0^\prime }^{\frac{\pi}{2}}\! \left[
\frac{1}{x_0^2}\tanh\frac{\hbar v_x x_0 }{4LT}\!-\!
\frac{4}{\pi^2} \tanh\frac{ \pi\hbar v_x }{8LT} \right]\!\tan
x_0\,
dx_0\\
+\int_0^{x_0^\prime} \frac{dx_0}{\tilde x_0}\tanh\frac{\hbar v_x
\tilde x_0}{4LT}\ .
\end{eqnarray*}
Here $\tilde x_0= \sqrt{x_0^2+ (2L\Gamma/\hbar v_x)^2
\cos^2(\phi/2)}$ and $L\Gamma/\hbar v_{2D}\ll x_0^\prime \ll 1$.
For low temperatures,
\[
B = \left\{ \begin{array}{lr}
\begin{displaystyle} \ln
\left(\frac{\hbar v_{2D}}{L\Gamma|\cos(\phi/2)|}\right)
\end{displaystyle} , & T\ll \Gamma |\cos(\phi/2)|\\
\begin{displaystyle} \ln \left(\frac{\hbar v_{2D}}{LT}\right)
\end{displaystyle} , &
\Gamma|\cos(\phi/2)| \ll T\ll \begin{displaystyle}\frac{\hbar
v_{2D}}{L}\end{displaystyle}
\end{array} \right.
\]
It is independent of $\theta$ within the logarithmic accuracy.
Performing the summation we find for $T\ll v_{2D}/L$
\begin{equation}
I=\frac{\pi}{eR_0}\left(\frac{L\Gamma^2}{\hbar v_{2D}}\right)B\sin
\phi \ . \label{current2}
\end{equation}
The current-phase relation is sinusoidal; this corresponds to the
limit of low transparency junctions $Z\gg 1$ realized in
short-lead contacts with $L\Gamma/\hbar v_{2D} \ll 1$. The
supercurrent increases with lowering the temperature and saturates
for $T\lesssim \Gamma$. For temperatures higher than $\hbar
v_{2D}/L$, the logarithm decreases to a value of order unity, and
our approximation breaks down. Note that the current-phase
relation, Eq.~(\ref{current2}), is similar to that obtained for a
double barrier SINIS structure with resonant transmission (see,
e.g., Ref.~\onlinecite{galaktionov}).

If the leads are very short, $L\ll \hbar v_{2D}/\Delta$, there
exists only one level, Eq. (\ref{E-bound}). The estimates show
that, within the logarithmic approximation, the current is
dominated by the spectrum close to the resonance, Eq.
(\ref{E-res}), and still has the form of Eq. (\ref{current2}).

\subsubsection{Long leads}

Here we again restrict ourselves to the limit $\Gamma \ll \Delta$.
For the contact with long leads, $L\Gamma/\hbar v_x \gg 1$ the
spectrum consists of one level with $\epsilon_0 <\Gamma$
satisfying Eqs. (\ref{E<}), (\ref{A<}) and a series of levels with
$\epsilon_n
>\Gamma$ satisfying Eqs. (\ref{E<}), (\ref{A>}).
Typical spectrum of bound states for a contact with long leads is
plotted in Fig.~\ref{fig-boundstates}(b).

For $\epsilon<\Gamma$ we find from Eqs. (\ref{E<}), (\ref{A<})
\begin{equation}
x  =\frac{(2L\Gamma/\hbar v_x) \sin (\phi/2)\sinh x} {\sqrt{
\sin^2\alpha^\prime +\sinh^2x}} \label{eq<}
\end{equation}
where $x=(2L\Gamma/\hbar v_x) \sqrt{1- \epsilon^2/\Gamma^2 } $. If
$L\Gamma/\hbar v_x \gg 1$ we have $\sinh x \gg 1$ thus $x =
(2L\Gamma/\hbar v_x) \sin (\phi/2)$, i.e., this energy state is as
in a ballistic junction with a gap $\Gamma$,
\[
\epsilon^2 =\Gamma^2 \cos^2(\phi/2)\ .
\]
This level gives the usual expression for the supercurrent as in a
ballistic contact\cite{Kulik78} with an induced gap $\Gamma$
\cite{AVolkov95},
\begin{equation}
I=\frac{\pi\Gamma \sin(\phi/2)}{eR_0}\tanh\frac{\Gamma \cos
(\phi/2)}{2T} \ . \label{sup-curr}
\end{equation}

One also needs to consider the contribution of levels with
energies above the induced gap. For $L\Gamma/\hbar v_x \gg 1$ Eqs.
(\ref{E<}) and (\ref{A>}) yield $x^\prime =\pi n +\delta x_n$
where $x^\prime =(2L\Gamma/\hbar v_x) \sqrt{\epsilon^2/\Gamma^2-1
} $ and
\begin{equation}
\pi n +\delta x_n  =b|\sin  (\delta x_n)|\ .
\end{equation}
Here we denote $ b^2= (2L\Gamma/\hbar v_x)^2 \sin^2 (\phi/2)/
\sin^2\alpha^\prime $. Expanding in $b^{-1}\ll 1$ we find $ \delta
x_n =\pm  \pi n/ b - \pi n/b^2$. For $\epsilon \gg \Gamma$ we have
\[
\sum_n \frac{\partial \epsilon_n}{\partial \phi}=- \frac{\hbar
v_x}{2L} \sum_n \left[\pi n\frac{\partial }{\partial \phi}
\frac{1}{ b^2}\right]\ .
\]
The terms in $\delta x$ proportional to $b^{-1}$ disappear. The
number of terms in the sum is $N\sim b$, being determined by the
condition $\delta x \lesssim 1$. Therefore, the sum over the
states with $\epsilon
>\Gamma$ is of the order of $ (\hbar v_x/2L)\ll \Gamma $, and can thus be
neglected. As a result, Eq. (\ref{sup-curr}) is the full
expression for supercurrent through a contact with long leads. In
this sense, the contact with long leads behaves as a contact
without normal reflection, $|r_N|=0$, when the bound states would
only exist for $\epsilon <\epsilon_g$. This limit was considered
in Ref.\cite{AVolkov95}.

\section{The I-V curve in a 2D/S junction} \label{sec-IV}

\begin{figure}[t]
\centering
\includegraphics[width=0.75\linewidth]{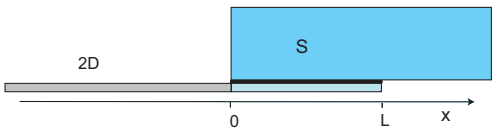}
\caption{Junction made of the superconducting electrode placed on
top of the 2D electron gas.}\label{fig-junction}
\end{figure}

The results obtained in the previous section can be used to
describe the transport in a 2D/S junction shown in
Fig.~\ref{fig-junction}. The junction consists of a semi-infinite
normal 2D layer a part of which (with a length L) is covered by a
bulk superconducting lead. The current through such contact can be
written in terms of reflection coefficients obtained in Sec.
\ref{subsec-reflection}. Using the results of Ref.~\cite{BTK82},
we have
\begin{eqnarray*}
I &=& \frac{1}{eR_0} \int _{-\infty}^\infty  d\epsilon \,
\int_{v_x>0} \frac{d \theta}{2}\, \frac{v_x}{v_{2D}}
\left[1-|r_N|^2+|r_A|^2\right]  \\
&&\times[f_0(\epsilon -eV)-f_0(\epsilon)]
\end{eqnarray*}
where $ f_0(\epsilon)=[e^{\epsilon/T}+1]^{-1} $ is the Fermi
function. For $T\ll \Gamma$ the differential conductance is
\begin{equation}
\frac{dI}{dV}= \frac{1}{R_0} \int_{0}^{\pi/2} \cos \theta
\left[1-|r_N|^2+|r_A|^2\right]\, d \theta \ , \label{NIScurrent1}
\end{equation}
where $\epsilon =eV$.

\subsection{Long leads. Andreev reflection} \label{subsec-Long}

Consider the case $L\to \infty$ such that the length is longer
than the electronic mean free path, $L \gg \ell $. In this limit
one can neglect the effect of the dead end and assume that $Z=0$.

For $Z=0$ we have zero normal reflection amplitude $r_N=0$ and the
Andreev reflection coefficient $ |r_A|^2= |v|^2/|u|^2 $ which is
independent of $p_y$. For $\epsilon >\Delta$ the Andreev
reflection coefficients becomes
\[
|r_A|^2= \left|\frac{\epsilon
\left(1+\frac{i\Gamma}{\sqrt{\epsilon^2 -\Delta^2}}\right) -
\sqrt{\epsilon^2\left( 1+\frac{2i\Gamma}{\sqrt{\epsilon^2
-\Delta^2}}\right)-\Gamma^2}}{\epsilon
\left(1+\frac{i\Gamma}{\sqrt{\epsilon^2 -\Delta^2}}\right) +
\sqrt{\epsilon^2\left( 1+\frac{2i\Gamma}{\sqrt{\epsilon^2
-\Delta^2}}\right)-\Gamma^2}}\right|
\]
For $\epsilon <\Delta$ we have
\[
|r_A|^2= \left|\frac{\epsilon
\left(1+\frac{\Gamma}{\sqrt{\Delta^2-\epsilon^2}}\right) -
\sqrt{\epsilon^2\left( 1+\frac{2
\Gamma}{\sqrt{\Delta^2-\epsilon^2}}\right)-\Gamma^2}}{\epsilon
\left(1+\frac{\Gamma}{\sqrt{\Delta^2-\epsilon^2}}\right) +
\sqrt{\epsilon^2\left(
1+\frac{2\Gamma}{\sqrt{\Delta^2-\epsilon^2}}\right)-\Gamma^2}}\right|
\]

\begin{figure}[b]
\centering
\includegraphics[width=0.8\linewidth]{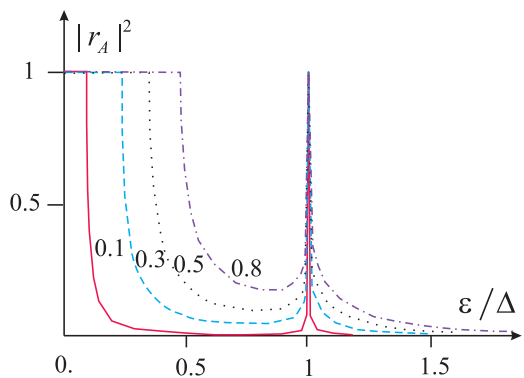}
\caption{Andreev reflection coefficient as a function of energy
for zero normal reflection, $Z=0$. Here
$x=\epsilon/|\Delta|$.}\label{fig-rA}
\end{figure}

If $\epsilon^2 <\epsilon_g^2$, one has $|u|=|v|$ and $|r_A|^2=1$.
If $\epsilon \to \Delta -0$ we also have $ |r_A|^2 \to 1 $. If
$\Gamma \ll \Delta$ then for $\epsilon_g \ll \epsilon \ll \Delta$
we have $ |r_A|^2= \Gamma^2/4\epsilon^2$. Since the Andreev
reflection does not depend on $\theta$, Eq. (\ref{NIScurrent1})
yields for $T\ll \Gamma$
\[
R_0 \frac{dI}{dV}= 1+|r_A|^2
\]
where $\epsilon=eV$. The Andreev reflection coefficient for a long
contact is shown in Fig. \ref{fig-rA}. The limit of a long contact
with zero normal reflection was considered in Ref.
\cite{Fagas-etal-05} using a direct-contact model.

\subsection{Finite-length leads. Oscillations of the resistance}

Here we discuss contacts with finite-length leads in the weak
coupling limit, $\Gamma \ll |\Delta|$. For $\epsilon<|\Delta|$
when quasiparticles cannot escape into the leads, Eq.
(\ref{norm1}) gives $ 1-|r_N|^2+|r_A|^2 =2 |r_A|^2 $. Equation
(\ref{NIScurrent1}) yields
\[
R_0\frac{dI}{dV}=2\int_{0}^{\pi/2} \cos\theta |r_A(\epsilon)|^2 \,
d \theta\ .
\]
In the low-voltage region, $\epsilon <\epsilon_g$, we have $u^*
=v$ while $Z$ is real. Therefore,
\[
|r_A|^2= \frac{4u^2v^2\sinh^2[2L\sqrt{\Gamma^2-\tilde \epsilon^2}
/\hbar v_x]}{4u^2v^2\cosh^2[2L\sqrt{\Gamma^2-\tilde \epsilon^2}
/\hbar v_x] -1}\ .
\]

In the weak coupling case $\epsilon _g=\Gamma$ and
\[
\eta_1=\epsilon \Gamma/\Delta\ , \; \eta_2 =\Gamma\ , \;
4u^2v^2=\Gamma^2/\epsilon^2\ , \; \tilde \epsilon = \epsilon\ .
\]
The differential conductance becomes
\[
R_0\frac{dI}{dV}= 2 \times \left\{
\begin{array}{lr}\begin{displaystyle}
\frac{4L^2}{\xi_{2D}^2}\ln \frac{\hbar v_{2D}}{L\Gamma}
\end{displaystyle} \ , & L\Gamma /\hbar v_{2D} \ll 1 \\
 1\ , & L\Gamma /\hbar v_{2D} \gg 1\end{array} \right.
\]
It does not depend on voltage. For short leads $ L\Gamma /v_{2D}
\ll 1$, the conductance is determined by the incident angles
$\theta$ close to $\pi/2$ where the Andreev reflection is of the
order of unity. For long leads $ L\Gamma /\hbar v_{2D} \gg 1$ the
Andreev reflection is complete as in the previous Section
\ref{subsec-Long}.

For energies $\epsilon_g \ll \epsilon < |\Delta|$, the Andreev
coefficient is
\[
|r_A|^2=\frac{2u^2v^2[1-\cos
[2(\delta_+-\delta_-)]]}{u^4+v^4-2u^2v^2\cos
[2(\delta_+-\delta_-)]}\ .
\]
If the energy is not very close to the superconducting gap,
$|\epsilon-\Delta| \gg \Gamma^2/\Delta$, we have $v^2 \ll u^2$,
while $u^2=1$. Therefore
\[
|r_A|^2=2v^2[1-\cos [2(\delta_+-\delta_-)]]
\]
and $ v^2(\epsilon)= \Gamma^2/4\epsilon ^2$. Therefore, for
$\epsilon_g \ll eV <\Delta$,
\[
R_0\frac{dI}{dV}= \frac{4L^2}{\xi_{2D}^2} \times \left\{\!\!
\begin{array}{lr} \begin{displaystyle} 2  \ln \frac{1}{\tilde V}
\end{displaystyle} , & \!\! \tilde V \ll 1 \\
\begin{displaystyle}\frac{1}{\tilde
V^2}\left[1+ \sqrt{\frac{\pi}{2\tilde V}}\sin\!\left(2\tilde V -
\frac{\pi}{4} \right)\right]\end{displaystyle} ,& \!\! \tilde V
\gg 1
\end{array}\right.
\]
where $\tilde V=2LeV/\hbar v_{2D}$. The conductance exhibits
oscillations as a function of the bias voltage due to the
geometrical quantization.

\begin{figure}[b]
\includegraphics[width=0.6\linewidth]{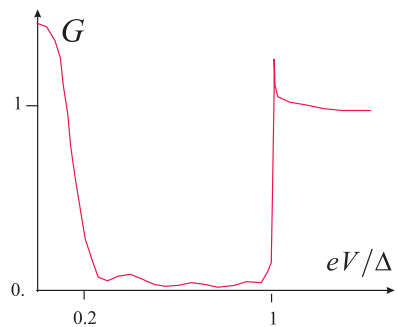}
\caption{Differential conductance $G=R_0(dI/dV)$ as a function of
$V/100\Delta$. Here $\Gamma/\Delta =0.1$ and $2L \Delta/\hbar
v_{2D}=10$.}\label{fig-dIdV}
\end{figure}

Consider now the energies $eV>\Delta\gg \Gamma$. Not very close to
the gap edge $\Delta$ the Andreev reflection is small, while
\[
|r_N|^2 = \exp [-2{\rm Im}\, (\delta_+-\delta_-)]=\exp
\left[-\frac{4L\Gamma}{\hbar
v_x}\frac{\epsilon}{\sqrt{\epsilon^2-\Delta^2}} \right]\ .
\]
For short contacts $L\Gamma /\hbar v_{2D}\ll 1$ we find
\begin{equation}
R_0\frac{dI}{dV}= \frac{2\pi L\Gamma}{\hbar
v_{2D}}\frac{\epsilon}{\sqrt{\epsilon^2-\Delta^2}}\ .
\label{G-tunnel}
\end{equation}
Region close to the gap edge, $|\epsilon-\Delta| \to 0$ needs a
special consideration. Here $u^2\to v^2$, so that the normal
reflection vanishes while the Andreev reflection grows up to
$|r_A|^2=1$, and the conductance has a sharp peak as in the case
of long contacts considered in Sec. \ref{subsec-Long}. The
differential conductance is plotted in Fig.~\ref{fig-dIdV}.

As was mentioned in Sec. \ref{subsec-reflection}, short leads are
equivalent to the limit of low transmission tunnel contact. The
differential conductance Eq. (\ref{G-tunnel}) at voltages
$eV>\Delta$ is proportional to the superconducting density of
states, as it is usually the case for tunnel contacts. However, in
addition to the peak at the superconducting gap edge, the
differential conductance has a low-energy peak which characterizes
the induced superconducting gap $\epsilon_g$ in the 2D layer. In
the voltage interval $\epsilon_g <eV< \Delta$, the conductance
oscillates as a function of voltage due to the geometrical
quantization in the regions where the leads overlap with the 2D
layer. The contacts with long leads, $L\Gamma /\hbar v_{2D}\gg 1$,
for $\epsilon>\Delta$ have an exponentially small normal
reflection due to almost complete escape of quasiparticles into
the superconductors, and thus their conductance coincides with
that obtained in Sec. \ref{subsec-Long}.

\section{Discussion} \label{sec-disc}

We formulate the approach which can be used for spatially
inhomogeneous and/or time-dependent problems associated with the
induced superconductivity in low dimensional electronic systems
including a 2D electron gas, graphene layer, etc. This approach is
based on the so-called Fano--Anderson model which describes the
decay of a resonance state coupled to a continuum
\cite{FanoAnderson}. We consider a 2D system placed in a contact
with a bulk superconductor and use the tunnel approximation for
coupling between the superconductor and the 2D electron layer in a
way similar to that used in Refs. \cite{Shiba,ArseevVolkov91} for
impurities in a superconductor.  We consider two particular
examples of junctions made of a ballistic 2D electron gas placed
under the superconducting electrodes of a finite length. For the
case of a short symmetric S/2D/S junction we find the bound states
localized in the junction and calculate the supercurrent as a
function of the lead length. Next we consider the 2D/S junction
and find the IV curve for various lead lengths. We show that the
differential conductance as a function of the bias voltage
exhibits peaks which correspond to the induced and bulk
superconducting gaps. The differential conductance also shows
oscillations as a function of voltage due to geometrical
quantization in the regions with the induced superconductivity.

\acknowledgements

We thank A. Ioselevich and S. Sharov for stimulating discussions.
This work was supported in part by the Academy of Finland, Centers
of excellence program 2006--2011, by the Russian Foundation for
Basic Research under grant 09-02-00573-a,  by Russian Agency of
Education under the Federal Program ``Scientific and educational
personnel of innovative Russia in 2009-2013'' and by the Program
``Quantum Physics of Condensed Matter'' of the Russian Academy of
Sciences.

\end{document}